\documentclass[journal]{IEEEtran}
\makeatletter

\usepackage{blindtext}
\usepackage[T1]{fontenc}
\usepackage{textcomp}
\usepackage{amsmath}
\usepackage{amssymb}
\usepackage{bm}
\usepackage{mdwmath}
\usepackage{multirow}
\usepackage{makecell}
\usepackage{comment}
\usepackage{cases}
\usepackage[short, c2]{optidef}
\usepackage{graphicx}
\graphicspath{{Figure/PDF/}{Biography/PDF/}}
\usepackage[dvipsnames]{xcolor}
\definecolor{myblue}{rgb}{0,0.4980,1} 
\definecolor{myred}{rgb}{0.8706,0.1608,0.0627} 

\usepackage[caption=false,font=footnotesize,subrefformat=parens]{subfig}

\usepackage[nosort,noadjust]{cite}
\usepackage{url}
\usepackage{tabularray}
\UseTblrLibrary{varwidth}
\UseTblrLibrary{diagbox}
\ExplSyntaxOn
\tl_const:Nn \c__tblr_valign_M_tl {M}
\tl_new:N    \g__tblr_cell_valign_sub_tl

\keys_define:nn { tblr-column }
  {
    M .meta:n = { valign = M },
  }

\cs_gset_protected:Npn \__tblr_get_cell_alignments:nn #1 #2
  {
    \group_begin:
    \tl_gset:Nx \g__tblr_cell_halign_tl
      { \__tblr_data_item:neen { cell } {#1} {#2} { halign } }
    \tl_set:Nx \l__tblr_v_tl
      { \__tblr_data_item:neen { cell } {#1} {#2} { valign } }
    \tl_case:NnF \l__tblr_v_tl
      {
        \c__tblr_valign_t_tl
          {
            \tl_gset:Nn \g__tblr_cell_valign_tl {m}
            \tl_gset:Nn \g__tblr_cell_middle_tl {t}
            \tl_gclear:N \g__tblr_cell_valign_sub_tl
          }
        \c__tblr_valign_m_tl
          {
            \tl_gset:Nn \g__tblr_cell_valign_tl {m}
            \tl_gset:Nn \g__tblr_cell_middle_tl {m}
            \tl_gclear:N \g__tblr_cell_valign_sub_tl
          }
        \c__tblr_valign_M_tl
          {
            \tl_gset:Nn \g__tblr_cell_valign_tl {m}
            \tl_gset:Nn \g__tblr_cell_valign_sub_tl {M}
            \tl_gset:Nn \g__tblr_cell_middle_tl {t}
          }
        \c__tblr_valign_b_tl
          {
            \tl_gset:Nn \g__tblr_cell_valign_tl {m}
            \tl_gset:Nn \g__tblr_cell_middle_tl {b}
            \tl_gclear:N \g__tblr_cell_valign_sub_tl
          }
      }
      {
        \tl_gset_eq:NN \g__tblr_cell_valign_tl \l__tblr_v_tl
        \tl_gclear:N \g__tblr_cell_middle_tl
      }
    \group_end:
  }

\cs_gset_protected:Npn \__tblr_build_cell_content:NN #1 #2
  {
    \hbox_set_to_wd:Nnn \l__tblr_a_box { \l__tblr_cell_wd_dim }
      {
        \tl_if_eq:NnTF \g__tblr_cell_halign_tl {j}
          { \__tblr_get_cell_text:nn {#1} {#2} \hfil }
          {
            \tl_if_eq:NnF \g__tblr_cell_halign_tl {l} { \hfil }
            \__tblr_get_cell_text:nn {#1} {#2}
            \tl_if_eq:NnF \g__tblr_cell_halign_tl {r} { \hfil }
          }
      }
    \vbox_set_to_ht:Nnn \l__tblr_b_box { \l__tblr_cell_ht_dim }
      {
        \tl_case:Nn \g__tblr_cell_valign_tl
          {
            \c__tblr_valign_m_tl
              {
                \vfil
                \bool_lazy_and:nnT
                  { \int_compare_p:nNn { \l__tblr_cell_rowspan_tl } < {2} }
                  { !\tl_if_eq_p:NN \c__tblr_valign_M_tl \g__tblr_cell_valign_sub_tl }
                  {
                    \box_set_ht:Nn \l__tblr_a_box
                      { \__tblr_data_item:nen { row } {#1} { @row-upper } }
                    \box_set_dp:Nn \l__tblr_a_box
                      { \__tblr_data_item:nen { row } {#1} { @row-lower } }
                  }
                \box_use:N \l__tblr_a_box
                \vfil
              }
            \c__tblr_valign_h_tl
              {
                \box_set_ht:Nn \l__tblr_a_box
                  { \__tblr_data_item:nen { row } {#1} { @row-head } }
                \box_use:N \l__tblr_a_box
                \vfil
              }
            \c__tblr_valign_f_tl
              {
                \vfil
                \int_compare:nNnTF { \l__tblr_cell_rowspan_tl } < {2}
                  {
                    \box_set_dp:Nn \l__tblr_a_box
                      { \__tblr_data_item:nen { row } {#1} { @row-foot } }
                  }
                  {
                    \box_set_dp:Nn \l__tblr_a_box
                      {
                        \__tblr_data_item:nen
                          { row }
                          { \int_eval:n { #1 + \l__tblr_cell_rowspan_tl - 1 } }
                          { @row-foot }
                      }
                  }
                \box_use:N \l__tblr_a_box
              }
          }
        \hrule height ~ 0pt 
      }
    \vbox_set_to_ht:Nnn \l__tblr_c_box
      { \l__tblr_row_ht_dim - \l__tblr_row_abovesep_dim }
      {
        \box_use:N \l__tblr_b_box
        \vss
      }
    \skip_horizontal:n { \l__tblr_x_tl }
    \box_use:N \l__tblr_c_box
    \skip_horizontal:n { \l__tblr_y_tl - \l__tblr_cell_wd_dim + \l__tblr_w_tl }
  }
\ExplSyntaxOff
\usepackage{pifont}
\UseTblrLibrary{counter}
\usepackage[linesnumbered,ruled]{algorithm2e}

\usepackage{hyperref}
\newcommand{\colorhypersetup}{\@ifpackageloaded{hyperref}{\hypersetup{%
	bookmarksopen=true,%
	bookmarksnumbered=true,%
	pdfpagemode={UseOutlines},
	pdfstartview={FitH},%
	colorlinks=true,%
	linkcolor={myred},%
	citecolor={orange}
}}{\empty}}
\newcommand{\blackhypersetup}{\@ifpackageloaded{hyperref}{\hypersetup{%
    colorlinks=true,%
    linkcolor=.,%
    citecolor=.,%
    filecolor=.,%
    urlcolor=.
}}{\empty}}
 \blackhypersetup

\usepackage{mfirstuc-english}
\MFUhyphentrue
\usepackage[version3,description,IEEEtran]{eacro}
\acsetup{%
    pages/display=all,%
    pages/seq/use=false,%
    pages/name=true,%
    pages/fill={\quad},%
    make-links=true%
}
\DeclareAcronym{ela}{
	short = ELA,
	long = experience level agreement}
\DeclareAcronym{bs}{
	short = BS,
	long = base station}
 \DeclareAcronym{es}{
	short = ES,
	long = edge server}
 \DeclareAcronym{mg}{
	short = MG,
	long = multicast group}
  \DeclareAcronym{msvs}{
	short = MSVS,
	long = multicast short video streaming}
 \DeclareAcronym{ai}{
	short = AI,
	long = artificial intelligence}
  \DeclareAcronym{dt}{
	short = DA,
	long = digital agent}
 \DeclareAcronym{andcm}{
	short = ADT4AM,
	long = AI-native network DT architecture for autonomous network management}
 \DeclareAcronym{udt}{
	short = UDT,
	long = user DT}
 \DeclareAcronym{idt}{
	short = IDT,
	long = infrastructure DT}
 \DeclareAcronym{sdt}{
	short = SDT,
	long = slice DT}
 \DeclareAcronym{nc}{
	short = NC,
	long = network controller}
\DeclareAcronym{cnn}{
	short = CNN,
	long = convolutional neural network}
\DeclareAcronym{rnn}{
	short = RNN,
	long = recurrent neural network}
 \DeclareAcronym{gan}{
	short = GAN,
	long = generative adversarial network}
  \DeclareAcronym{nlp}{
	short = NLP,
	long = natural language processing}
 \DeclareAcronym{bleu}{
	short = BLEU,
	long = bilingual evaluation understudy}
 \DeclareAcronym{rouge}{
	short = ROUGE,
	long = recall-oriented understudy for gisting evaluation}
 \DeclareAcronym{gnn}{
	short = GNN,
	long = graph neural network}
  \DeclareAcronym{drl}{
	short = DRL,
	long = deep reinforcement learning}
 \DeclareAcronym{llm}{
	short = LLM,
	long = large language model}
  \DeclareAcronym{ap}{
	short = AP,
	long = access point}
   \DeclareAcronym{iot}{
	short = IoT,
	long = Internet of things}
 \DeclareAcronym{qoe}{
	short = QoE,
	long = quality of experience}
 \DeclareAcronym{qos}{
	short = QoS,
	long = quality of service}
 \DeclareAcronym{lstm}{
	short = LSTM,
	long = long short-term memory}
 \DeclareAcronym{ddqn}{
	short = DDQN,
	long = double deep Q network}
  \DeclareAcronym{cs}{
	short = CS,
	long = cloud server}
 \DeclareAcronym{svm}{
	short = SVM,
	long = support vector machine}
 \DeclareAcronym{knn}{
	short = K-NN,
	long = K-nearest neighbors}
  \DeclareAcronym{gai}{
	short = GAI,
	long = generative AI}
      \DeclareAcronym{ar}{
	short = AR,
	long = augmented reality}
    \DeclareAcronym{sla}{
	short = SLA,
	long = service level agreement}

\newcommand{\upperroman}[1]{\uppercase\expandafter{\romannumeral#1}}

\newcommand{\myunit}[1]{%
	\ifmmode
		\mathrm{#1}
	\else
		$ \mathrm{#1} $
	\fi}
\newcommand{\murm}{%
	\ifmmode
		\text{\textmu}
	\else
		\textmu
	\fi}

\newcommand{\MYnewpage}{%
	\ifCLASSOPTIONonecolumn
		\ifCLASSOPTIONjournal
			\typeout{The onecolumn journal mode.}
			\newpage
		\fi
	\fi}

\newlength{\mysinglefigwidth}
\newlength{\mymultifigwidth}
\ifCLASSOPTIONonecolumn
	\AtBeginDocument{\setlength{\mysinglefigwidth}{0.7\linewidth}}
\else
	\AtBeginDocument{\setlength{\mysinglefigwidth}{\linewidth}}
\fi
\ifCLASSOPTIONonecolumn
	\AtBeginDocument{\setlength{\mymultifigwidth}{0.5\linewidth}}
\else
	\AtBeginDocument{\setlength{\mymultifigwidth}{\linewidth}}
\fi

\makeatother
\begin{document}
\ifCLASSOPTIONonecolumn
    \typeout{The onecolumn mode.}
    \title{\LARGE Title}
    \author{Author~1,~\IEEEmembership{Member,~IEEE}, and~Author~2,~\IEEEmembership{Fellow,~IEEE}
        
    }
\else
    \typeout{The twocolumn mode.}
    \title{Revolutionizing QoE-Driven Network Management with Digital Agents in 6G}
     \author{Xuemin (Sherman) Shen,~\IEEEmembership{Fellow,~IEEE}, Xinyu Huang,~\IEEEmembership{Student Member,~IEEE}, Jianzhe Xue,~\IEEEmembership{Student Member,~IEEE}, Conghao Zhou,~\IEEEmembership{Member,~IEEE}, Xiufang Shi,~\IEEEmembership{Member,~IEEE}, Weihua Zhuang,~\IEEEmembership{Fellow,~IEEE}
    
     \thanks{Xuemin (Sherman) Shen, Xinyu Huang, Jianzhe Xue, Conghao Zhou, and Weihua Zhuang are with the Department of Electrical and Computer Engineering, University of
	Waterloo, Waterloo, ON N2L 3G1, Canada (E-mail: \{sshen, x357huan, j59xue, c89zhou, wzhuang\}@uwaterloo.ca). Xiufang Shi is with the College of Information Engineering, Zhejiang University of Technology, Hangzhou, 310023, China (E-mail: xiufangshi@zjut.edu.cn).
     }}
    
\fi

\ifCLASSOPTIONonecolumn
	\typeout{The onecolumn mode.}
\else
	\typeout{The twocolumn mode.}
\fi

\maketitle

\ifCLASSOPTIONonecolumn
	\typeout{The onecolumn mode.}
	\vspace*{-50pt}
\else
	\typeout{The twocolumn mode.}
\fi
\begin{abstract}
In this article, we present a digital agent (DA)-assisted network management framework for future sixth generation (6G) networks considering user quality of experience (QoE). A novel QoE metric is defined by incorporating the impact of user behavioral dynamics and environmental complexity on quality of service (QoS). A two-level DA architecture is proposed to assist the QoE-driven network slicing and orchestration. Three potential solutions are presented from the perspectives of DA data collection, resource scheduling, and DA deployment. A case study demonstrates that the proposed framework can effectively improve user QoE compared with benchmark schemes.

\end{abstract}

\ifCLASSOPTIONonecolumn
	\typeout{The onecolumn mode.}
	\vspace*{-10pt}
\else
	\typeout{The twocolumn mode.}
\fi
\begin{IEEEkeywords}
Digital agent, network slicing and orchestration, quality of experience (QoE), experience level agreement (ELA).
\end{IEEEkeywords}

\IEEEpeerreviewmaketitle

\MYnewpage


\section{Introduction}
\label{sec:Introduction}

As fifth-generation (5G) networks approach maturity and widespread deployment, both industry and academia are turning their attention to the sixth generation (6G) wireless networks. The 6G is anticipated to support an unprecedented diversity of services with significantly increased service demands, driving the evolution towards experience-centric networking~\cite{wang2023road,holi}. \Ac{qoe}, as a subjective performance metric, can reflect the satisfaction level of users on networking services~\cite{yang20226g}. Different users usually have differentiated QoE model structures and parameters~\cite{gao2020personalized}. Therefore, QoE-driven network management is essential to meet the diverse and personalized requirements of users in 6G networks.

The \ac{dt} concept emerges as a promising avenue to advance QoE-driven network management in 6G networks. A \ac{dt} is an intelligent proxy of a physical entity on the network side, which is expected to perform efficient behavior emulation and data analytics for customized network management. Unlike traditional digital twins in manufacturing~\cite{grieves2014digital}, which focuses on simulation and monitoring, \acp{dt} in communication networks enhance network management capabilities by modeling network behaviors and analyzing intrinsic network features. {Existing automation capabilities in communication networks, such as analytics services and decision-making functions, typically operate under the assumption that information about network behavior is readily available within these functions~\cite{3gpp28915}. To alleviate this issue, DAs are anticipated to provide new information on how the network behaves or will behave, enabling a deeper understanding and modeling of network dynamics.} This information is particularly valuable in use cases involving the measurement of user satisfaction with network services. By integrating predictive analytics and QoE modeling, DAs are expected to contribute to a user-specific QoE-driven network management framework. 


However, achieving the user-specific QoE-driven network management framework faces technical challenges. Firstly, traditional QoE models are learned offline and do not account for the impact of real-time user behaviors and environmental factors. Therefore, how to efficiently construct and update comprehensive QoE models for individual users on the network side is worth studying. Secondly, due to the diversity in QoE models and the heterogeneity of user statuses, purely data-driven or model-driven methods often struggle to quickly and accurately allocate network resources to individual users to achieve satisfactory QoE. Therefore, it is essential to develop an efficient and tailored network orchestration strategy. Thirdly, existing network slicing strategies mainly focus on resource demand estimation based on \ac{qos} rather than \ac{qoe}, which can sometimes lead to a mismatch between actual resource demands and reserved network resources. Therefore, developing an adaptive network slicing strategy from the perspective of QoE is essential to address this issue.

\begin{table*}[t]
\centering
\caption{Influencing factors of user QoE}

\begin{tabular}{l|lc|lc|c|c}
\hline\hline
\multicolumn{1}{c|}{\multirow{2}{*}[-0.35cm]{ User QoE}} & \multicolumn{1}{c|}{QoS}                                                                                     & Emotion                                                                                             & \multicolumn{1}{c|}{Device}                                                                                  & Scenario                                                                      & Application & Environment                                                                                                               \\ \cline{2-7} 
\multicolumn{1}{c|}{}                          & \multicolumn{1}{l|}{\begin{tabular}[c]{@{}l@{}}Dealy, throughput, \\packet loss,\\ jitter, etc.\end{tabular}} & \begin{tabular}[c]{@{}l@{}}Joy, calm, sadness, \\ anger, surprise, \\fear, disgust, etc.\end{tabular} & \multicolumn{1}{l|}{\begin{tabular}[c]{@{}l@{}}Smartphone, tablet, \\ computer, gaming \\console, etc.\end{tabular}} & \begin{tabular}[c]{@{}l@{}}Home, workplace, \\ mobile, public, etc.\end{tabular} & \begin{tabular}[c]{@{}l@{}}Social media, entertainment, \\ office, e-commerce, etc.\end{tabular} & \begin{tabular}[c]{@{}l@{}}Complexity, \\ dynamics \end{tabular}                          \\ \hline

\end{tabular}
\label{qoe}
\end{table*}

{In this article, we propose a novel \ac{dt}-assisted network management framework to enhance user \ac{qoe} in future 6G networks. The key contributions are summarized as follows:}
    \begin{itemize}
        \item {Comprehensive QoE modeling: We introduce a new QoE metric that integrates dynamic user behaviors and environmental complexity into traditional QoS-oriented approaches to provide an accurate and personalized measure of user satisfaction.}
        \item {Two-level \ac{dt} architecture: We propose a hierarchical \ac{dt} architecture to facilitate QoE-driven network orchestration and slicing. Level-one \acp{dt} handle user-specific QoE modeling, resource demand prediction, and tailored network orchestration, while level-two \acp{dt} abstract resource demand distribution and make adaptive slice adjustments.}
        \item {Performance evaluation: We present a case study of a video streaming scenario to demonstrate the effectiveness of the proposed framework. The experiment results show that our \ac{dt}-assisted framework can notably improve user QoE as compared to benchmark schemes.}
    \end{itemize}

{
The remainder of this paper is organized as follows. Section~II reviews the preliminaries of QoE-driven network management. Section III describes the proposed two-level DA-assisted framework. A case study evaluating the framework's performance is presented in Section IV. Section~V discusses research challenges and potential solutions, and Section VI concludes this work.}

\section{Preliminaries of QoE-Driven Network Management}

\subsection{User QoE}

User \ac{qoe} is a multifaceted metric for evaluating communication network performance from the user's perspective, encapsulating the overall satisfaction when interacting with network services~\cite{kao2023qoe}. As summarized in Table~\ref{qoe}, user \ac{qoe} is influenced by a combination of objective and subjective factors:

\begin{itemize}
    \item \textbf{\Ac{qos} factors}: {Network performance metrics such as delay and packet loss directly impact the responsiveness and reliability of services. For example, maintaining network jitter below 30 ms and latency under 10 ms is critical for real-time interactive services to avoid QoE reduction.}

    \item \textbf{Emotion}: User emotional states, including joy, calm, sadness, anger, surprise, fear, and disgust, can influence their perception of service quality. For instance, a frustrated user may perceive network issues more negatively than a content user.

    \item \textbf{Device}: The type of devices, e.g., smartphone, tablet, computer, and gaming console, affects the user's experience due to variations in processing power, display quality, and user interfaces.

    \item \textbf{Scenario}: The context in which the service is accessed, such as at home, in the workplace, on the move, or in public spaces, can alter network conditions and user expectations, thereby influencing \ac{qoe}.

    \item \textbf{Application}: Different applications, including social media, entertainment, office productivity, and e-commerce, have varying performance requirements and user expectations, which play a significant role in the perceived \ac{qoe}.

    \item \textbf{Environment}: The complexity and dynamics of the environment affect user \ac{qoe}. In a highly complex and dynamic environment, users usually require higher \ac{qos} to maintain service performance. For example, \ac{ar} users in a complex environment usually require more computing resources for video rendering to guarantee high-definition video playback.
    Users in a high-speed train usually need more data caching in their devices to avoid service interruption. Therefore, proper responding to the environmental context is essential for enhancing user \ac{qoe}.
    
\end{itemize}

By integrating these factors, network operators can better understand user diverse requirements and make tailored network management strategies to enhance user \ac{qoe}.

\subsection{Network Resource Management for QoE Enhancement}

The advent of 6G networks brings unprecedented opportunities and challenges in network resource management aimed at enhancing user \ac{qoe}. Different from the previous generations, 6G networks are expected to support ultra-high data rates, massive connectivity, ultra-reliable and low-latency communications (URLLC), intelligent networking capabilities, and advanced sensing functions \cite{wang2023road}. Effective management of communication, computing, caching, and sensing resources becomes imperative to meet these demands. This subsection explores how 6G networks can enhance user QoE through resource management in these four critical areas.

\subsubsection{Communication resource management}

Efficient communication resource management is foundational for optimizing QoE in 6G networks. By leveraging new spectrum bands, such as in terahertz frequencies, and advanced technologies like massive multiple-input multiple-output (MIMO), 6G networks will achieve ultra-high data rates essential for bandwidth-intensive applications. This enhancement directly impacts user QoE by enabling services like holographic communications and immersive virtual reality, providing seamless and high-quality experiences. Intelligent spectrum management utilizing AI-driven algorithms dynamically allocates frequency bands based on real-time demand, network conditions, and QoE requirements of users \cite{luo2022resource}. This approach ensures efficient spectrum utilization and interference reduction, leading to user satisfaction. Moreover, URLLC is used to meet stringent latency and reliability needs, which is critical for applications involving both user interactions and device communications, thus enhancing overall QoE. 

\begin{figure*}[!t]
    \centering
    \includegraphics[width=1.58\mysinglefigwidth]{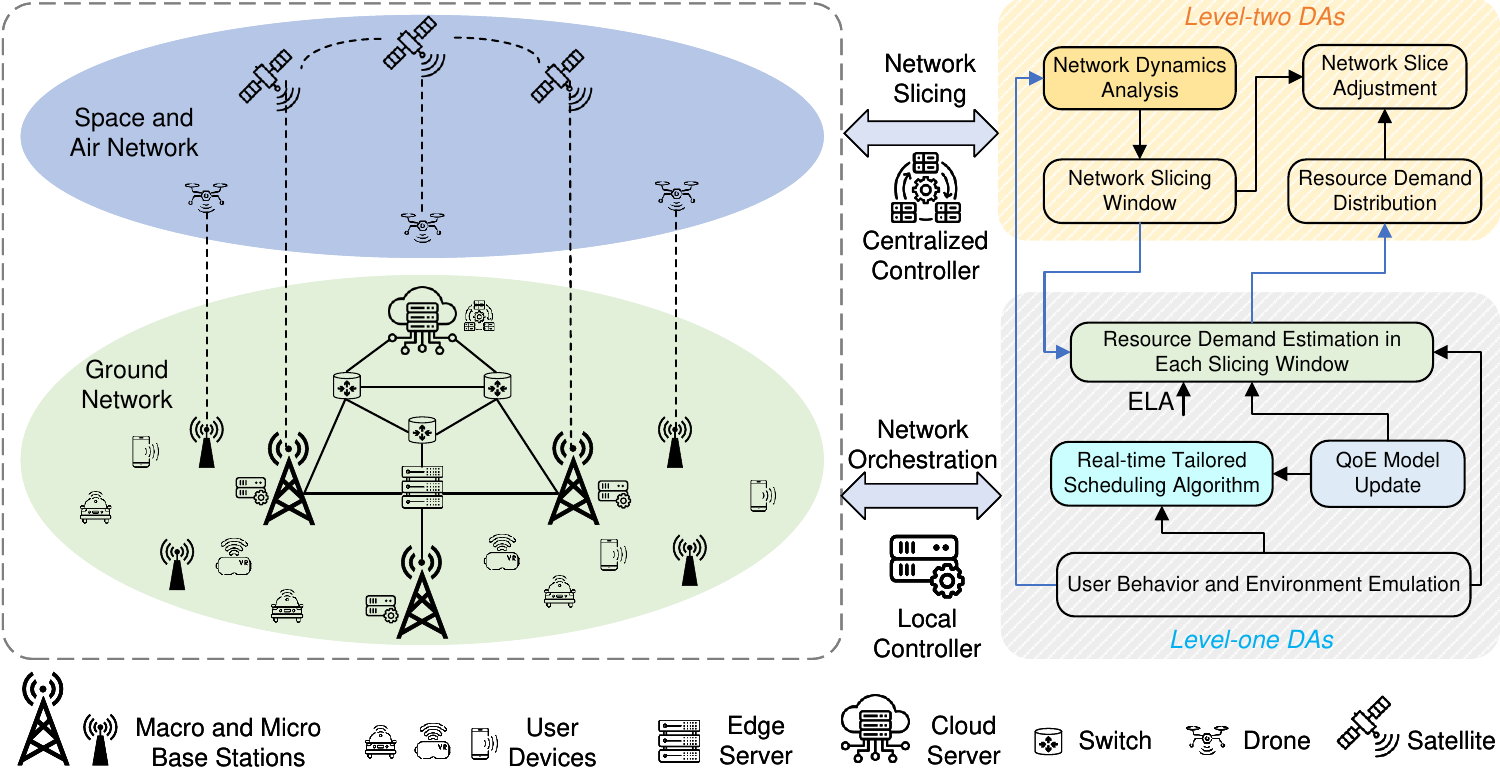}
    \caption{DA-assisted network management framework.}
    \label{frame}
\end{figure*}
\subsubsection{Computing resource management}

Computing resource management in 6G networks will be characterized by the integration of edge intelligence, distributed computing, and AI capabilities to meet the demands of compute-intensive applications. Pervasive edge computing with native AI brings computational resources and decision-making close to users, which can efficiently orchestrate computation tasks between devices, edge servers, and cloud servers~\cite{kai2020collaborative}. This enhancement improves user QoE by providing responsiveness in real-time applications, such as augmented reality (AR) and gaming. Distributed computing and federated learning allow collaborative processing and model training without centralized data aggregation \cite{sannara2021federated}. This approach improves QoE by enabling devices to learn from shared models while preserving privacy and reducing communication overhead. 

\subsubsection{Caching resource management}

Caching strategies in 6G networks can be enhanced through intelligent content distribution and storage management, aiming to reduce latency and improve data availability for users. Semantic and proactive caching stores semantically significant content and data at the network edge based on predictive analytics~\cite{zhang2020psac}. For users, this ensures that relevant information is readily available, enhancing QoE by reducing access delays and buffering times. Utilizing distributed ledger technologies like blockchain for caching ensures data integrity, security, and transparency, enhancing user trust and ensuring that devices operate with reliable and secure data. Edge caching with AI optimization uses algorithms to optimize caching decisions by learning content popularity patterns, user preferences, and device data needs, reducing backhaul traffic and access latency.

\subsubsection{Sensing resource management}
6G networks integrate advanced sensing capabilities with communication functions, enabling the networks to perceive and interpret the physical environment~\cite{wang2023road}. Efficient sensing resource management involves coordinating sensory data collection, processing, and dissemination to support context-aware services and intelligent decision-making~\cite{shaoirs}. For users, enhanced sensing means improved experiences in services such as immersive AR and personalized applications that adapt to the user environment and context, thereby enhancing QoE.

\section{A Digital Agent-Assisted Network Management Framework}

\subsection{Framework Overview}

As shown in Fig.~\ref{frame}, we propose a DA-assisted network management framework to improve user QoE. The left part is the physical networks, consisting of space and air networks and ground networks. Specifically, the ground networks utilize a cloud-edge-end architecture to distribute computing and storage resources close to users, which can optimize data processing and reduce latency for user QoE enhancement. The space and air networks are utilized to extend network coverage to remote and underserved regions, provide high-speed data backhaul, and ensure seamless connectivity across diverse geographic areas. The right part is the proposed two-level DAs. {Particularly, level-one DAs, mainly deployed at the edge server, are utilized to emulate user behaviors and environment changes and to maintain user QoE models for experience-centric network orchestration. On the other hand, level-two DAs, mainly deployed at the cloud server, abstract resource demand distribution from level-one DAs to facilitate experience-centric network slicing.} {Local controllers at base stations or edge nodes handle real-time network orchestration, while centralized controllers in data centers manage long-term network slicing. In the proposed framework, URLLC can be used for DA data collection, while intelligent networking can be used to solve complex resource management problems.}

\subsection{Digital Agent}
The knowledge of how the network behaves or will behave is the new value provided by DAs that can understand and model the behaviors of the network. {Network behavior is a broad concept that includes both user behavior and operational behavior of a network infrastructure.} Therefore, we propose a two-level \ac{dt} architecture to analyze user behaviors and contextual information to construct user-specific QoE models for experience-centric network management.

Level-one DA is a virtual representation of an individual user, which records its ID, service requests, quality of service (QoS), \ac{ela}, and contextual information. {Compared with \ac{sla}, ELA serves a threshold for the constructed QoE model to obtain a satisfactory network service experience~\cite{kao2023qoe}.} Specifically, user ID and service requests are utilized to differentiate QoE models for different network services. The QoS, including service delay, throughput, jitter, and packet loss, are used to measure objective service quality. User contextual information, such as mobility, emotion, and environment, are analyzed to emulate user behaviors and environment dynamics. {Mobility reflects a user's location trajectory through the network, while emotion, such as service satisfaction level, influences user behavior, such as engagement level.} Based on the information, a comprehensive QoE model can be established. In the network orchestration process, users can be clustered into different groups based on their QoE model structures, where tailored resource scheduling algorithms can be designed to solve specific QoE optimization problems. Furthermore, the established QoE models should be integrated with \acp{ela} as well as user behaviors and environment dynamics to estimate user resource demands in a network slicing window determined by level-two DAs.

Level-two DAs aggregate user information from level-one DAs to abstract two distilled information, i.e., resource demand distribution and network dynamics. The abstracted information of network dynamics should be utilized to adjust the network slicing window, which is fed back to the resource demand estimation module in level-one DAs. Compared with traditional QoS-based resource demand estimation, user QoE models in level-one DAs are used to estimate which and how many network resources are required to improve user QoE, which can better reflect user actual demands. The abstracted resource demand distribution, network slicing window, and current virtualized network resources are inputted to network slicing optimization problems, which are solved by data-driven, model-driven, or combined methods to adjust network slices.

{In a scenario with high user density, edge resources can be dynamically allocated, and workload can be redistributed across multiple edge servers coordinated by level-two DAs. To adapt to changing mobility patterns, predictive models in level-one DAs can enable proactive resource allocation.}

\subsection{Level-One DA-Assisted Network Orchestration}
To facilitate experience-centric network orchestration, level-one DAs are utilized to construct and update comprehensive QoE models for individual users, estimate resource demands for level-two DAs, and make tailored network orchestration strategies.

\subsubsection{Comprehensive QoE model construction and update}
In this subsection, we will introduce how to construct a comprehensive QoE model in a level-one DA and an adaptive QoE model update method.

\begin{figure}[!t]
    \centering
    \includegraphics[width=0.95\mysinglefigwidth]{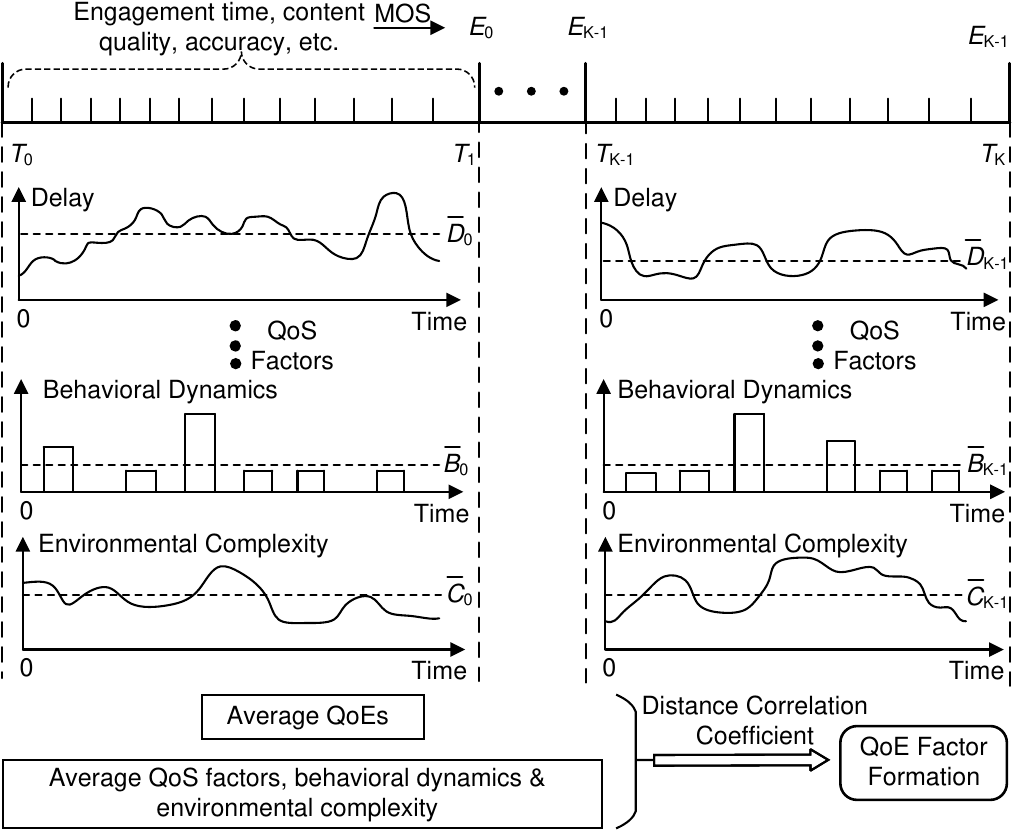}
    \caption{QoE factor formation.}
    \label{attr}
\end{figure}
\begin{itemize}
    \item QoE factor formation: As shown in Fig.~\ref{attr}, to determine the relationship between QoE and factors such as QoS, behavioral dynamics, and environmental complexity, users’ feedback data over a specific period are stored and analyzed in level-one DAs. The data such as engagement time, content quality, and accuracy are utilized to estimate the mean opinion score (MOS), which can provide a quantitative measure of overall QoE experienced by users~\cite{agarwal2022qoe}. Subsequently, the average values of QoS factors, behavioral dynamics, and environmental complexity during the analyzed period are integrated with the QoE scores to constitute a multivariate vector. This vector is then input into the distance correlation coefficient (DCC) analysis to identify which factors significantly correlate with QoE. The DCC, ranging from 0 (no correlation) to 1 (perfect correlation), quantifies the strength of the relationship~\cite{edelmann2021relationships}. Finally, the mathematical relationship is modeled by inputting QoE and its correlated influencing factors into numerical fitting methods.

    \item User behavior and environment emulation:
    Level-one DAs, primarily deployed on edge computing servers, leverage advanced predictive algorithms such as finite state machines (FSMs) and recurrent neural networks (RNNs) to emulate user behavior effectively. These models utilize contextual information about users to simulate actions and responses in dynamic network environments. Additionally, sophisticated scene generation algorithms, like generative adversarial networks (GANs), are used by level-one DAs for environmental modeling and simulating operational status. Monte Carlo simulations are used to predict changes in complex and varying network environments. By emulating both user behavior and environmental changes, user QoE models can reflect user realistic resource demands.

    \item Comprehensive QoE model establishment:
    To accurately reflect the realistic satisfaction of users with network services, a comprehensive QoE model is essential. This model integrates three key components: QoS, behavioral dynamics, and environmental complexity. {As shown in Eq.~\eqref{qoe_1}, a comprehensive QoE model for each user is constructed by integrating the impact of user behavior and environment on the objective QoS metric.} 
    \begin{equation}\label{qoe_1}
    	E=S\times I(B, C),
    \end{equation}
    {where $E$ and $S$ represent QoE and QoS, respectively. Function $I(B,C)$ is the impact function, which quantifies how behavioral dynamics $B$ and environmental complexity $C$ influence QoS. The structure and parameters of impact function can be obtained through data fitting methods. Eq.~\eqref{qoe_1} is developed for customized resource management. For example, if a user utilizing the AR service remains relatively stationary and the surrounding environment is simple, the QoS requirement can be more relaxed, leading to lower resource demand. By using Eq.~\eqref{qoe_1}, user contextual information is integrated into the QoE model to enable customized resource management.}
    
    \item Adaptive QoE model update:
    The adaptive QoE model update mechanism is essential for experience-centric network management. Since level-one DAs can emulate future user behaviors and environment, the changes in emulated user behavior patterns and environmental complexity are assessed to determine whether to trigger the QoE model update. The update focuses on two key components: QoE model factors and their corresponding weights. The relevance of various factors is evaluated through the DCC analysis to identify which factors are currently most influential. This ensures that the QoE model remains sensitive to the most pertinent aspects of the user experience. Concurrently, the weights of these factors are optimized using data fitting techniques, aligning the model more closely with the observed trends and predictions derived from DA emulations. 
    
\end{itemize}

{The constructed QoE models can not only capture the impact of environmental factors but also indirectly account for user emotion to enable tailored resource allocation strategies.}

\subsubsection{QoE-based resource demand prediction}

\begin{figure}[!t]
    \centering
    \includegraphics[width=0.9\mysinglefigwidth]{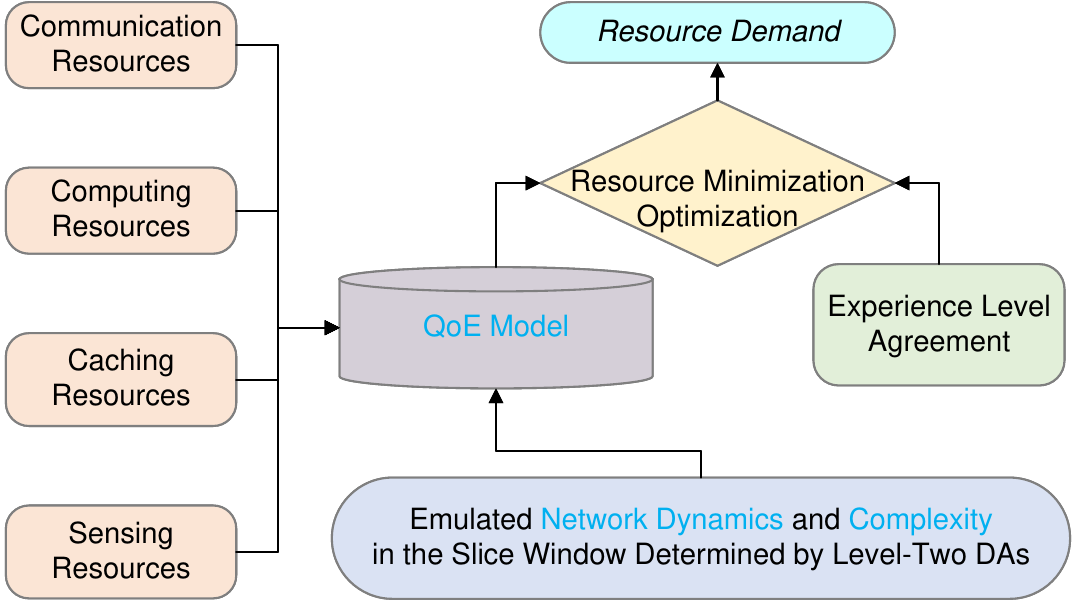}
    \caption{QoE-based resource demand prediction procedure.}
    \label{demand}
\end{figure}

As shown in Fig.~\ref{demand}, we propose a QoE-based resource demand procedure. Here, the resource demands of users are estimated in a slice window determined by level-two DAs, since \ac{ela} reflect users' long-term experience requirements. The network resources including communication, computing, caching, and sensing resources are integrated with emulated network dynamics and complexity in a slice window to constitute each user-specific QoE model. The constructed QoE model and \ac{ela} are then input to the network resource minimization problem, which can be solved by model-based approaches such as convex optimization and dynamic programming, and data-based approaches such as \ac{drl} and \ac{gai}. Therefore, level-one DAs can flexibly provide each user's estimated resource demand for level-two DAs to conduct further feature abstraction.

\subsubsection{Tailored network orchestration approaches}

\begin{figure}[!t]
    \centering
    \includegraphics[width=\mysinglefigwidth]{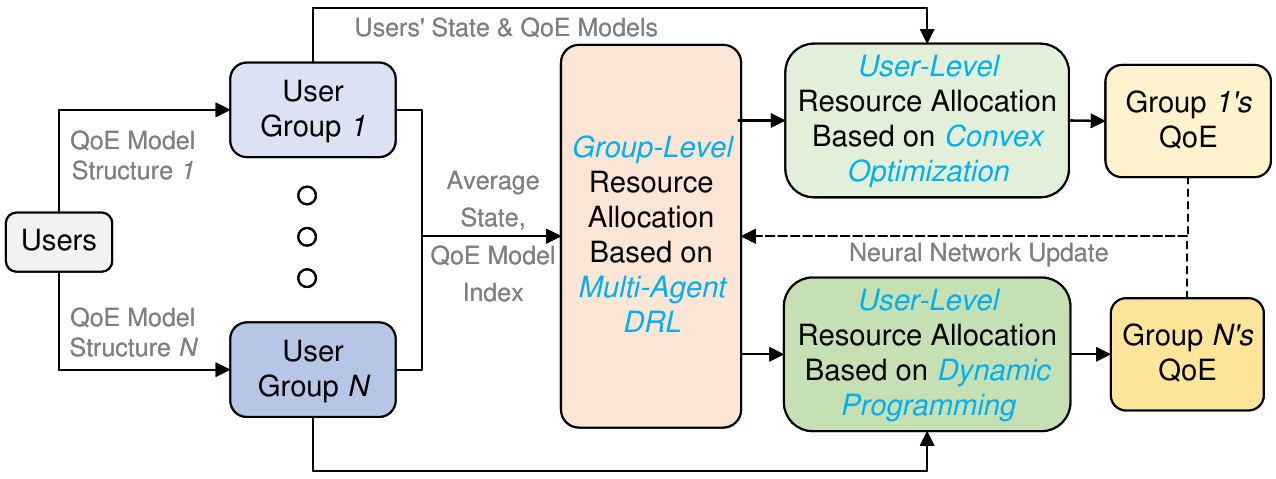}
    \caption{Tailored network resource orchestration process.}
    \label{orch}
\end{figure}

As shown in Fig.~\ref{orch}, we propose a tailored network resource orchestration process. Specifically, users are first clustered into different groups based on the QoE model structures. The users in one group have the same QoE model structures but their model parameters can be different. Then, the average group-level state, such as buffer length, computing load, and content quality, is integrated with the QoE model index, as an input of group-level resource allocation module. This module can adopt a multi-agent \ac{drl} algorithm to conduct resource allocation strategies for each group. Next, the group-level resource allocation strategies are then combined with user state and QoE models, as an input of user-level resource allocation module. Based on the differences in optimization problems of various groups, this module can be customized. For instance, if the optimization problem can be transformed into a convex problem, a convex optimization method can be adopted to achieve the optimal QoE. If the optimization problem can be transformed into a Knapsack problem, a dynamic programming method can be utilized to achieve satisfactory QoE. Finally, the groups' QoE is fed back to the group-level resource allocation module for neural network update in the model training stage. 
\subsection{Level-Two DA-Assisted Network Slicing}
To facilitate experience-centric network slicing, level-two DAs are utilized to adjust the resource demand estimation window for level-one DAs, abstract resource demand distribution from level-one DAs, and make adaptive network slice reconfiguration.

\subsubsection{Resource demand distribution abstraction}
The resource demand distribution abstraction relies on user resource demand estimation in a network slice window. The resource demand has two unique labels, i.e., service type and location. The resource demands of users using the same service in the same base station coverage are aggregated to constitute the resource demand distribution. If the network resources are sufficient, network resources can be directly sliced based on the resource demand distribution. Otherwise, network resources need to be preferentially sliced to user groups that can achieve the highest QoE gain (The QoE increases with the same network resources).

\subsubsection{Adaptive slice reconfiguration}
To adapt to network dynamics, network slices require adaptive adjustments, including adaptive slice window and game-based slice adjustment.
\begin{itemize}
    \item Adaptive slice window:    
    As illustrated in Fig.~\ref{frame}, the emulated user behaviors and environmental information in level-one DAs are transferred to the network dynamics analysis module in level-two DAs. Network dynamics are then analyzed based on variations in user behaviors and environmental data. The analyzed network dynamics are discretized into multiple stages, with each stage corresponding to a specific slice window size. Higher network dynamics generally correlate with shorter slice window sizes. The mapping between network dynamics and slice window size can also be fine-tuned based on QoE feedback.
    
    \item Game-based slice adjustment:
    In game-based slice adjustment, multiple user groups with diverse \ac{ela} compete for limited network resources to enhance their individual QoE. Each user group acts as a rational player in a non-cooperative game, aiming to optimize its own reserved network resources. The competition is modeled using game theory, where utility functions quantify the QoE improvements relative to the reserved network resources. Players strategize to maximize their utilities by adjusting their slice configurations of communication, computing, caching, and sensing resources. The interaction among user groups leads to an equilibrium state—typically a Nash Equilibrium—where no group can independently improve its QoE without adversely affecting others. By incorporating algorithms like potential games and iterative best-response dynamics, the network can dynamically adjust slices in response to real-time user behavior and environmental changes, which can ensure an efficient and fair distribution of network resources.

\end{itemize}

\section{Case Study}
We present a case study to evaluate the performance of proposed DA-assisted network management framework in video streaming scenarios. The simulation region selects the University of Waterloo campus, where two base stations with $700~\myunit{MHz}$ band (n28: $703-803$~\myunit{MHz}) and a connected edge server with $10~\myunit{GCycles/s}$ computing capacity are deployed. The number of users in the simulation region is set as $k\in\{16, 18, 20, 22, 24\}$. Each user moves along a prescribed path with a speed ranging from $2\sim 40~\myunit{Km/h}$. The channel path losses between users and base stations are generated based on the PropagationModel at Matlab. The transmitted video sequences are sampled from the YouTube 8M dataset (https://research.google.com/youtube8m/index.html). The video bitrate ranges from $500~\myunit{Kbps}$ to $3~\myunit{Mbps}$. Users' video requests are modeled as a Poisson distribution with an arrival rate of $\lambda=6$. Users’ MOS are generated based on three principles, rebuffer time-based probability distribution $\mathcal{N}(5-0.4R, 8; 1, 5)$, quality-based one $\mathcal{N}(1+4Q,1;1,5)$, and combined one $\mathcal{N}(1+4Q-0.4R, 0.8; 1,5)$, where $R$ and $Q$ represent rebuffer time and video quality, respectively. Users’ behavioral dynamics and environmental complexity range from $1$ to $2$, where the front relies on swipe frequencies, and the latter has a negative correlation relationship with velocities. Users' ELAs are uniformly sampled with the range of $[3,5]$, and their QoE models are fitted by using the lsqcurvefit function at Matlab. {In the level-one DAs, the resource demand distribution is analyzed by integrating user QoE models and ELAs. The tailored scheduling algorithm selects a four-layer branch dueling Q network (BDQN)~\cite{tavakoli2018action} for group-level resource allocation and a sequential least squares quadratic programming (SLSQP) method for user-level resource allocation.} 
{In the level-two DAs, network dynamics is split into five parts, each of which corresponds to a slicing window size $W\in\{3, 6, 9, 12, 15\}~\myunit{min}$. The network slicing problem regarding communication and computing resource reservation is transformed into a convex optimization problem, solved by a low-complexity greedy algorithm. The two-level DA framework can adapt to various network scales.}

We compare our DA-assisted network management framework with three benchmark schemes, i.e., without DA (w/o DA) scheme, purely \ac{drl}-based level-one DA (PDRL-L1) scheme, and heuristic \ac{sla}-based level-two DA (HSLA-L2) scheme. The w/o DA scheme utilizes general QoE models for resource demand estimation, round-robin scheduling methods for real-time resource allocation, and greedy-based network slicing methods. The PDRL-L1 scheme adopts a five-layer BDQN algorithm to substitute the data-model-driven scheduling algorithm in the level-one DA, while the HSLA-L2 scheme selects QoS models and SLA to estimate resource demand distribution for network slice adjustment in the level-two DA.

\begin{figure}[!t]
	\centering
	\subfloat[The cumulative distribution function (CDF) of ELA achievable ratio.]{
		\includegraphics[width=0.4\textwidth]{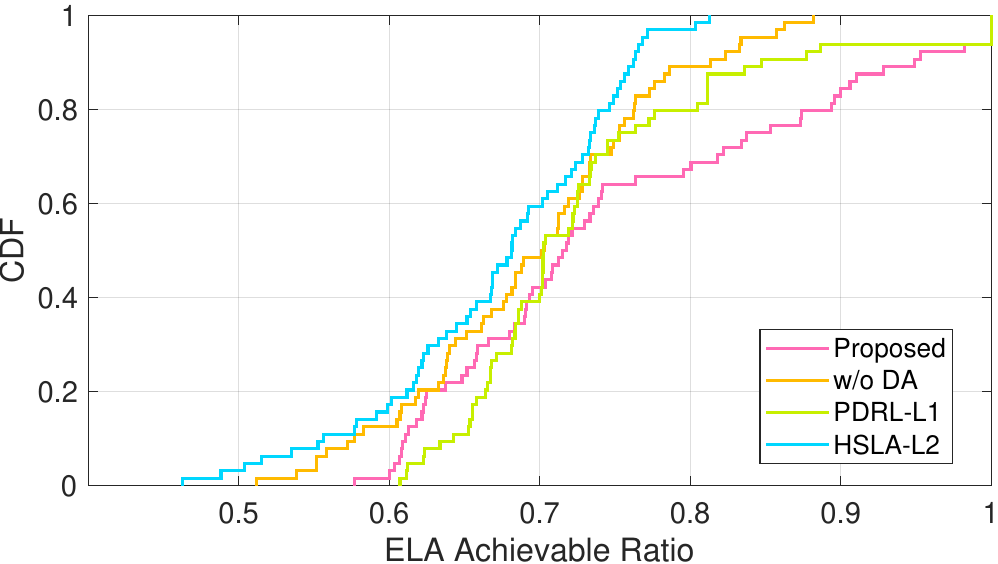}\label{5a}}
	\\
	\subfloat[The box plot comparison of real-time QoE.]{
		\includegraphics[width=0.39\textwidth]{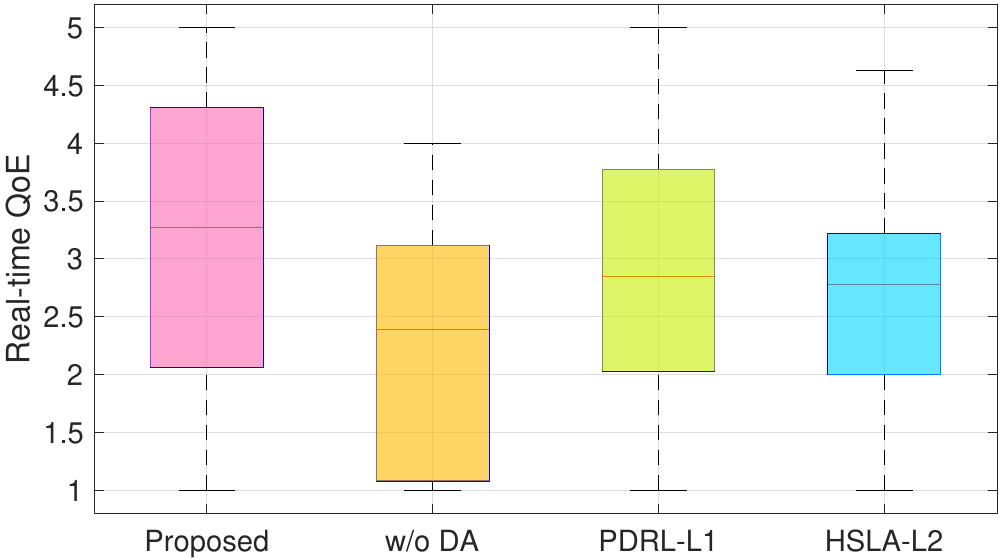}\label{5b}}
  \caption{ELA achievable ratio and real-time QoE comparison.}
	\label{fig:performance}
\end{figure}

As shown in Fig.~\subref*{5a}, we compare the CDF curves of ELA achievable ratio, i.e., the proportion of user QoE over ELA in each slicing window. It can be observed that the proposed framework can achieve the highest ELA achievable ratios in most percentile ranges. However, in the mid-range ($0.6-0.7$), the PDRL-L1 scheme slightly outperforms the proposed framework due to its focus on user fairness, despite achieving lower ELAs overall. The w/o DA scheme shows moderate performance but lags behind the proposed and PDRL-L1 schemes, which demonstrate the importance of DA-based optimization. Although the HSLA-L2 scheme exhibits the poorest performance, it validates the effectiveness of QoE models and ELA in resource management. {As shown in Fig.~\subref*{5b}, the proposed framework demonstrates the highest median QoE. Its interquartile range (IQR) is not the narrowest because users have diverse ELAs and our objective is to ensure that each user’s QoE surpasses their ELAs.} The w/o DA scheme shows the lowest median QoE and a wide IQR, because the round-robin scheduling method ignores users' diversified resource demands, which makes real-time resource allocation less targeted in improving user QoE.

\section{Research Challenges \& Potential Solutions}
Although the proposed framework can facilitate experience-centric network orchestration and slicing to improve user QoE, some challenges still exist that need to be solved.

\subsection{Efficient Multi-Modal Data Collection and Generation}

To construct and maintain level-one and level-two DAs, it is imperative to collect multi-modal data such as user behaviors, network performance metrics, and environmental contextual information. However, these data are inherently heterogeneous, encompassing various formats and update intervals, leading to high communication overhead in traditional network settings. To alleviate this issue, semantic communication can be employed to reduce the amount of data transmitted. By extracting and transmitting only the semantic essence, the communication load is significantly decreased while preserving the information necessary for DA maintenance. Furthermore, generating these multi-modal data in real-time introduces considerable computational overhead, even though DAs are deployed on edge and cloud servers. To address this challenge, lightweight data generation models such as knowledge distillation can be utilized to reduce computational costs. 

\subsection{Flexible Resource Scheduling Algorithm Selection}


The proposed tailored resource scheduling approach is a two-layer architecture, where the bottom layer adopts multi-agent reinforcement learning for group-level resource allocation, and the topper layer consists of customized algorithms dedicated to user-level resource allocation. However, due to the diversity of user demands and the dynamic nature of user states, finding suitable resource allocation algorithms for individual users is relatively challenging. To achieve flexibility in the second-layer algorithm, we can input the current optimization problems and user states into a meta-learning model or an automated machine learning (AutoML) model. These technologies are capable of rapidly searching for and generating the most appropriate optimization algorithms by learning from prior optimization experiences and adapting to new scenarios.

\subsection{Scalable Two-level DA Deployment and Migration}

Level-one DAs are typically deployed on edge nodes close to users to facilitate low-latency and real-time interactions, while level-two DAs are usually hosted on central nodes to leverage powerful computational resources for extensive data processing. However, due to the diversity of terminal services and varying user demands, selecting appropriate nodes for deploying these two-level DAs is crucial for enhancing processing speed and reducing latency. An effective deployment solution involves a dynamic node selection mechanism that considers factors such as service requirements, network conditions, and node capabilities. By analyzing these factors, the system can assign level-one and level-two DAs to optimal nodes within a hierarchical edge-cloud architecture, thereby optimizing resource utilization and performance. 

\section{Conclusion \& Future Directions}

In this article, we have proposed a novel DA-assisted network management framework to improve user QoE. Two-level DAs interact with each other to update QoE models, make tailored network orchestration strategies, and adaptively adjust network slices. Three research challenges from the perspectives of DA data collection and generation cost, network scheduling algorithm selection, and DA deployment and migration are explored and given potential solutions. 

For future research, more efforts should be directed to DA-assisted closed-loop network management, efficient interaction between DAs and physical networks, and accurate performance evaluation of DA system.

%
\bibliographystyle{IEEEtran}
\bibliography{IEEEabrv,Ref}
%
%

\end{document}